\documentclass{cargese}\usepackage{graphicx}

\let\footnote\savefootnote
\let\footnotetext\savefootnotetext 
\let\footnoterule\savefootnoterule 
\normallatexbib

\newcommand{\un}{\underline}
\renewcommand{\d}{\mathrm{d}}
\newcommand{\cov}{\mathrm{cov}}

\renewcommand{\theequation}{\arabic{equation}}

\begin{document}

\articletitle[Spacetime diffeomorphisms and the geodesic approximation]
{Spacetime diffeomorphisms\\and the geodesic approximation}


\author{J\"urg K\"appeli}



\affil{Spinoza Institute 
and Institute for Theoretical Physics
\\
University of Utrecht, Leuvenlaan 4, 3584 CE Utrecht, The Netherlands}    

\email{kaeppeli@phys.uu.nl}


\begin{abstract}
  We present a spacetime diffeomorphism invariant formulation of the geodesic
  approximation to soliton dynamics.
\end{abstract}

The geodesic approximation to the dynamics of solitons has been developed in
various contexts, and for some cases, a general formula for the metric on the
moduli space was given \cite{Harvey:1991hq,Gauntlett:1993sh}.  Such an
expression is lacking for the case of gravitation, though the moduli space
metric is known, {\em e.g.}, for extremal Reissner-Nordstr\"om black holes
\cite{Gibbons:cp,Ferrell:1987gf}. We are interested in understanding the
moduli space geometry of more complicated black hole solutions
\cite{LopesCardoso:2000qm}.  This note is based on \cite{deWit:2002xz}, where
important elements of a gauge and diffeomorphism invariant formulation of the
moduli space approximation are presented.

Consider Yang-Mills theory in $n$ spacetime dimensions, minimally coupled to a
scalar field and to a gravitational background,
\begin{equation}
\label{YM-action}
  S =
 \int \mathrm{d}^n x\;\sqrt{|g|}\;
\mathrm{Tr} \Big[ -\frac{1}{4} 
 F_{\mu\nu}\, F^{\mu\nu}- \frac{1}{2}
 {D}_\mu\phi \, {D}^\mu\phi - V(\phi)\Big]\,.
\end{equation}
The starting point of the geodesic approximation is a space of static or
stationary solutions of this theory parameterized by $N$ continuous
integration constants $X^a$, so-called collective coordinates.  In general,
the stationary solutions are subject to a class of residual gauge
transformations and diffeomorphisms that preserve stationarity,
\begin{equation}
  \mathcal{L}_{k} \xi = 0\,,\quad  \mathcal{L}_{k} \Lambda = 0\,,
\end{equation}
where $k$ is the timelike Killing vector, and $\xi$ and $\Lambda$ parameterize
diffeomorphisms and gauge transformations, respectively.  This degeneracy has
to be modded out when extracting the correct moduli space description. The
parameters $\xi$ and $\Lambda$ also depend on the collective coordinates
$X^a$, as inequivalent solutions may be subject to different transformations.
The residual spacetime diffeomorphisms are supplemented by reparametrizations
of the moduli space, which do not carry any spacetime dependence. One is
therefore dealing with an extended geometry and new connection components must
be introduced in order to define parallel transport on the bundles over the
extended base space.\footnote{Indices $\mu,\nu,\ldots$ denote spacetime
  indices, $a,b,\ldots$ moduli space indices. The $\Omega,\Sigma\,\ldots$ run
  over both spacetime and moduli space.  Tangent space indices are
  underlined.} The gauge connection is extended by $A_a \, \mathrm{d} X^a$,
while the extensions due to diffeomorphism invariance are encoded in the
enlarged vielbein,
\begin{equation}\
\label{vielbein}
  E_{\un \Omega}{}^{\Sigma} = \left(\begin{array}{cc} e_{\un \mu}{}^\nu 
  & 0 \\ e_{\un a}{}^\nu & e_{\un a}{}^b \end{array}\right)\,,
\end{equation} 
where $e_{\un \mu}{}^\nu$ is the spacetime vielbein, and $ e_{\un a}{}^b$ is a
spacetime independent reference frame for moduli space.  The new connection
fields parameterize the degeneracy with which moduli space translations, $X^a
\rightarrow X^a + \delta X^a$, are represented on the fields. The covariant
variation,
\begin{equation}\label{eq:covdef}
  \delta_{\cov}(\zeta) = {\cal L}_{\zeta} + \delta_{\mathrm{gauge}}(\zeta\cdot A)\,,
\end{equation}
contains a diffeomorphism with parameter $\zeta^{\Omega} = -\delta X^{\un a}
e_{\un a}{}^{\Omega}$, accompanied by a sum over all gauge transformations
with $\zeta^{\Omega}$ contracted with the corresponding gauge field as parameter.  The
(stationary) covariant variations of the scalar and the gauge are given by
\begin{equation}\label{eq:covfields}
  \delta_\cov \phi = \delta X^{\un a} D(A)_{\un a} \phi\,,\quad \delta_{\cov}
  A_\Omega = \delta X^{\un a}F(A)_{\un a \Omega}\,.
\end{equation}
The covariant variation of the vielbein involves the torsion tensor
$R(\Gamma)$,
\begin{equation}\label{eq:covvielbein}
\delta_{\cov} e_{\un \Omega}{}^{\Sigma} = -\delta X^{\un a}\, R(\Gamma)_{\un
  a\,\un \Omega}{}^{\Sigma} - D(\Gamma)_{\un \Omega} ( \delta X^{\un a}e_{\un a}{}^{\Sigma}) \,.
\end{equation}
The dynamics in the geodesic approximation is defined by the time flow vector
field $k$ and a time function $\tau$, subject to $k\cdot \partial\,\tau= 1$,
which parameterizes the flow on the integral curves of $k$. The collective
coordinates are promoted to point particles $X^a\rightarrow X^a(\tau)$. All
gauge transformations and diffeomorphisms now carry an implicit time
dependence through the collective coordinates. Using $k$ we can project
tensors onto directions along $k$ and onto directions along the
$(n-1)$-dimensional spacelike hypersurfaces perpendicular to $k$. The geodesic
lift is effected by adding covariant variations to derivatives along $k$,
while leaving the space components unchanged. This guarantees that all
configurations continue to extremize the potential such that the particle
motion takes place in the moduli space of solutions. The geodesic lift (which
we denote by a prime) for the covariant
derivative of the scalar field reads
\begin{equation}
  ({D_\Omega\phi})' 
= D_\Omega\phi + \frac{k_{\Omega}}{k^2}\, k\cdot \partial X^{\un a} D_{\un a}
  \phi\,.
\end{equation} 
When projected, this indeed yields
\begin{equation}
  (k\cdot{D\phi})' = k\cdot D \phi +\dot X^{\un a} D_{\un a}
  \phi\,,\quad  {}^{\perp}({D_\Omega\phi})' = {{}^{\perp}D_\Omega\phi}\,,
\end{equation}
where $\perp$ denotes the part perpendicular to $k$, and we used $ \dot
X(\tau)= k\cdot \partial X({\tau})$ for $k\cdot \partial\, \tau = 1$. Since
the covariant derivatives contain the gauge connection, one finds the
analogous structure for the gauge fields,
\begin{equation}
 A_\Omega' = A_\Omega + k^{-2}
k_\Omega \dot X^{\un a} A_{\un a}\,. 
\end{equation}
The result for the field strength is
\begin{equation}
  F(A)_{\Omega\Lambda}' =F(A)_{\Omega\Lambda} +  2{k^{-2}}k_{[\Omega} \,
  \dot  X^{\un a} F(A)_{\un a \Lambda]} \,.
\end{equation}
The defining conditions of the covariant variations $\delta_{\cov}'$ in the
time dependent situation are derived from enforcing generalized Leibniz rules such as
\begin{equation}\label{eq:Leibniz}
  \delta_{\cov}' (D(A) \phi)' =  D(A)' (\delta_{\cov}'\phi) -
  [\delta_{\cov}' A', \phi ]\,,
\end{equation}
which are the same as for the underlying field theory. Analogous relations
must hold for the variations of the various field strengths. They are essential in
order to identify the covariant field variations $\delta_{\rm cov}'$ with the
variations associated with the moduli action principle. For the case at hand 
one can achieve this by adding compensating covariant variations,
\begin{equation}\label{eq:comp}
  \delta_{\cov} ' (D(A)_{\Omega}\phi)' = \delta_{\cov}  (D(A)_{\Omega}\phi)' -
  \delta_{\cov} (k^{-2}k_{\Omega} \dot X^{\un a} e_{\un a}{}^{\Omega})
  D(A)_{\Omega} \phi \,.    
\end{equation}
Calculations at this point are rather subtle and further scrutiny is required
to establish the validity and the generalization of (\ref{eq:comp}). Using
this definition one indeed recovers (\ref{eq:Leibniz}), where $\delta_{\cov}'
A_{\Omega}' = \delta X^{\un a} F(A)_{\un a\Omega}'$ is calculated along the
same lines. Note that (\ref{eq:covfields}) is respected in the geodesic lift.
This last relation is a result of the modified transformation properties of
the gauge fields in the time dependent situation, $\delta_{\Lambda} A_{\Omega}' =
(D(A)_\Omega + k^{-2} k_\Omega\, \dot X^{\un a} D_{\un a}(A) )\Lambda$.  It is
important that the transformation properties with respect to diffeomorphisms
remain unaltered in the geodesic lift. This guarantees the invariance of the
action.

Usually, it is possible to work with a torsion free connection $\Gamma$. In the
vielbein formulation, this is achieved by imposing an irreducible set of
torsion constraints which are solved in terms of the spin connection. For the
extended geometry we cannot impose the vanishing of all torsion components,
since the local Lorentz transformations make up just a subgroup of
$\mathrm{SO}(n,1)\times\mathrm{ SO}(N)$. In our simple setting this has no
effect on the moduli space geometry. Dropping the constant potential terms,
the action (\ref{YM-action}) in the geodesic approximation is given by
\begin{equation}
  S[X(\tau)] = \int \d \tau\, \left(\frac{1}{2} G_{\un a\,\un b} \dot X^{\un a}
  \dot X^{\un b} + J_{\un a} \dot X^{\un b} \right)\,,\\
\end{equation}
with the moduli space metric and current defined by
\begin{eqnarray}
  G_{\un a\,\un b} &=& \int \d V^{\perp} \mathrm{Tr} \Big[ {F}_{\mu \un
  a} \, h^{\mu\nu}\,  {F}_{\nu\un b} + D_{\un a} \phi D_{\un b} \phi
  \Big]\,,\\
  J_{\un a} & = & \int  \d V^{\perp} \mathrm{Tr} \Big[ k^{\mu} {F}_{\mu\rho}\,
  h^{\rho\sigma}\,{ F}_{\un a \sigma}  + (k\cdot D\phi) D_{\un a} \phi\Big]\,.
\end{eqnarray}
Here, $\d V^{\perp}$ is the $(n-1)$-dimensional volume element of the induced
metric $h_{\mu\nu}= g_{\mu\nu} - k^{-2} k_{\mu} k_{\nu}$ on the spacelike
hypersurface, 
$\d V^{\perp} = \frac{\sqrt{h}}{|k|(n-1)!} k^{\sigma} \varepsilon_{\sigma
  \mu_2\cdots \mu_n} \d x^{\mu_2} \cdots \d x^{\mu_n}$, and $h^{\mu\nu} =
g^{\mu\nu} - k^{-2} k^{\mu} k^{\nu}$ is its hypersurface inverse.

For standard field theory actions, imposing the field equations for the
auxiliary connection $A_a$ and vielbein components $e_{\un a}{}^\nu$ gives
rise to a set of covariant constraint equations which can be solved for these
auxiliary fields. Their solutions are subsequently reinserted into the
expression for the moduli metric. These constraint equations reflect the
underlying invariances of the field theory.


\begin{chapthebibliography}{99}


\bibitem{Harvey:1991hq} J.~A.~Harvey and A.~Strominger,
Commun.\ Math.\ Phys.\  { 151} (1993) 221.

\bibitem{Gauntlett:1993sh}
J.~P.~Gauntlett,
Nucl.\ Phys.\ B { 411} (1994) 443.


\bibitem{Gibbons:cp}
G.~W.~Gibbons and P.~J.~Ruback,
Phys.\ Rev.\ Lett.\  { 57} (1986) 1492.

\bibitem{Ferrell:1987gf}
R.~C.~Ferrell and D.~M.~Eardley,
Phys.\ Rev.\ Lett.\  { 59} (1987) 1617.

\bibitem{LopesCardoso:2000qm}
G.~Lopes Cardoso, B.~de Wit, J.~K\"appeli and T.~Mohaupt,
JHEP { 0012} (2000) 019


\bibitem{deWit:2002xz}
B.~de Wit and J.~K\"appeli,
arXiv:hep-th/0211228.

\end{chapthebibliography}
\end{document}